\documentstyle[12pt,epsf,cite]{article}
 1
 1
 1

\begin{document}
\begin{titlepage}

\def\baselinestretch{1.2}
\topmargin     -0.25in
\vspace*{\fill}
\begin{center}
{\large \bf Semi-leptonic four-fermion final states in polarised $\gamma \gamma$ reactions: \\
Exact results {\it vs} an improved narrow width approximation
}
\vspace*{0.5cm}

\begin{tabular}[t]{c}
 
{\bf M.~Baillargeon$^{1}$,  G.~B\'elanger$^{2}$  and F.~Boudjema$^{2}$ }
 \\
\\
\\
{\it 1.  Grupo Te\'orico de Altas Energias, Instituto Superior T\'ecnico}\\
{\it Edif\'{\i}cio Ci\^encia (F\'{\i}sica)
P-1096 Lisboa Codex, Portugal}\\
{\it 2. Laboratoire de Physique Th\'eorique} 
EN{\large S}{\Large L}{\large A}PP
\footnote{URA 14-36 du CNRS, associ\'ee \`a l'E.N.S de Lyon et \`a
l'Universit\'e de Savoie.}\\
{\it Chemin de Bellevue, B.P. 110, F-74941 Annecy-le-Vieux, Cedex, France.}

\end{tabular}
\end{center}

\centerline{ {\bf Abstract} }
\baselineskip=14pt
\noindent
 {\small   We calculate the process $\gamma \gamma \rightarrow e^- \bar{\nu}_e u \bar{d}$  
with the full set of Feynman diagrams. We consider different implementations of 
the $W$ width and  different combinations of the photon polarisations. The results of an improved 
narrow width approximation based on $\gamma \gamma \rightarrow W^+ W^-$ with full spin-correlations that takes into account 
different angular cuts as well as cuts on the 
invariant masses are compared to those of the complete calculation. }
\vspace*{\fill}
\vspace*{0.1cm}
\rightline{ENSLAPP-A-635/97}
\rightline{FISIST/2-97/CFIF}
\rightline{January 1997}
\end{titlepage}
\baselineskip=18pt


\newcommand{\be}{\begin{equation}}
\newcommand{\beq}{\begin{equation}}
\newcommand{\eeq}{\end{equation}}
\newcommand{\ee}{\end{equation}}
\newcommand{\beqn}{\begin{eqnarray}}
\newcommand{\eeqn}{\end{eqnarray}}
\newcommand{\bea}{\begin{eqnarray}}
\newcommand{\ena}{\end{eqnarray}} 
\newcommand{\ra}{\rightarrow}
\newcommand{\su}{$ SU(2) \times U(1)\,$}
\newcommand{\gag}{$\gamma \gamma$ }
\newcommand{\gam}{\gamma \gamma }
\newcommand{\np}{Nucl.\,Phys.\,}
\newcommand{\pl}{Phys.\,Lett.\,}
\newcommand{\pr}{Phys.\,Rev.\,}
\newcommand{\prl}{Phys.\,Rev.\,Lett.\,}
\newcommand{\prep}{Phys.\,Rep.\,}
\newcommand{\zp}{Z.\,Phys.\,}
\newcommand{\sovjnp}{{\em Sov.\ J.\ Nucl.\ Phys.\ }}
\newcommand{\nuclinst}{{\em Nucl.\ Instrum.\ Meth.\ }}
\newcommand{\annp}{{\em Ann.\ Phys.\ }}
\newcommand{\intjmp}{{\em Int.\ J.\ of Mod.\  Phys.\ }}
 
\newcommand{\eps}{\epsilon}
\newcommand{\mw}{M_{W}}
\newcommand{\mww}{M_{W}^{2}}
\newcommand{\mwmw}{M_{W}^{2}}
\newcommand{\mhmh}{M_{H}^2}
\newcommand{\mz}{M_{Z}}
\newcommand{\mzz}{M_{Z}^{2}}

\newcommand{\lra}{\leftrightarrow}
\newcommand{\tr}{{\rm Tr}}
\def\ls1{{\not l}_1} 
\newcommand{\cms}{centre-of-mass\hspace*{.1cm}}

\newcommand{\dkg}{\Delta \kappa_{\gamma}}
\newcommand{\dkz}{\Delta \kappa_{Z}}
\newcommand{\dz}{\delta_{Z}}
\newcommand{\dgz}{\Delta g^{1}_{Z}}
\newcommand{\dgzt}{$\Delta g^{1}_{Z}\;$}
\newcommand{\la}{\lambda}
\newcommand{\lag}{\lambda_{\gamma}}
\newcommand{\laz}{\lambda_{Z}}
\newcommand{\lnl}{L_{9L}}
\newcommand{\lnr}{L_{9R}}
\newcommand{\lt}{L_{10}}
\newcommand{\lu}{L_{1}}
\newcommand{\ld}{L_{2}}
\newcommand{\cw}{\cos\theta_W}
\newcommand{\sw}{\sin\theta_W}
\newcommand{\tw}{\tan\theta_W}

\newcommand{\epm}{$e^{+} e^{-}\;$}
\newcommand{\epemt}{$e^{+} e^{-}\;$}
\newcommand{\epem}{e^{+} e^{-}\;}
\newcommand{\ememt}{$e^{-} e^{-}\;$}
\newcommand{\emem}{e^{-} e^{-}\;}
\newcommand{\eeww}{e^{+} e^{-} \ra W^+ W^- \;}
\newcommand{\eewwt}{$e^{+} e^{-} \ra W^+ W^- \;$}
\newcommand{\epemww}{e^{+} e^{-} \ra W^+ W^- }
\newcommand{\epemwwt}{$e^{+} e^{-} \ra W^+ W^- \;$}
\newcommand{\eennhht}{$e^{+} e^{-} \ra \nu_e \bar \nu_e HH\;$}
\newcommand{\eennhh}{e^{+} e^{-} \ra \nu_e \bar \nu_e HH\;}
\newcommand{\ppwg}{p p \ra W \gamma}
\newcommand{\wwhh}{W^+ W^- \ra HH\;}
\newcommand{\wwhht}{$W^+ W^- \ra HH\;$}
\newcommand{\ppwz}{pp \ra W Z}
\newcommand{\ppwgt}{$p p \ra W \gamma \;$}
\newcommand{\ppwzt}{$pp \ra W Z \;$}
\newcommand{\gamgamt}{$\gamma \gamma \;$}
\newcommand{\gamgam}{\gamma \gamma \;}
\newcommand{\egamt}{$e \gamma \;$}
\newcommand{\egam}{e \gamma \;}
\newcommand{\gamgamwwt}{$\gamma \gamma \ra W^+ W^- \;$}
\newcommand{\gamgamwwht}{$\gamma \gamma \ra W^+ W^- H \;$}
\newcommand{\gamgamwwh}{\gamma \gamma \ra W^+ W^- H \;}
\newcommand{\gamgamwwhht}{$\gamma \gamma \ra W^+ W^- H H\;$}
\newcommand{\gamgamwwhh}{\gamma \gamma \ra W^+ W^- H H\;}
\newcommand{\ggww}{\gamma \gamma \ra W^+ W^-}
\newcommand{\ggwwt}{$\gamma \gamma \ra W^+ W^- \;$}
\newcommand{\ggwwht}{$\gamma \gamma \ra W^+ W^- H \;$}
\newcommand{\ggwwh}{\gamma \gamma \ra W^+ W^- H \;}
\newcommand{\ggwwhht}{$\gamma \gamma \ra W^+ W^- H H\;$}
\newcommand{\ggwwhh}{\gamma \gamma \ra W^+ W^- H H\;}
\newcommand{\ggwwz}{\gamma \gamma \ra W^+ W^- Z\;}
\newcommand{\ggwwzt}{$\gamma \gamma \ra W^+ W^- Z\;$}
\def\smx{{\cal{S}} {\cal{M}}\;}

\newcommand{\ptu}{p_{1\bot}}
\newcommand{\vecptu}{\vec{p}_{1\bot}}
\newcommand{\ptd}{p_{2\bot}}
\newcommand{\vecptd}{\vec{p}_{2\bot}}
\newcommand{\ie}{{\em i.e.}}
\newcommand{\cm}{{{\cal M}}}
\newcommand{\cl}{{{\cal L}}}
\newcommand{\cd}{{{\cal D}}}
\newcommand{\cv}{{{\cal V}}}
\def\slashc{c\kern -.400em {/}}
\def\slashL{L\kern -.450em {/}}
\def\slashcl{\cl\kern -.600em {/}}
\def\Ww{{\mbox{\boldmath $W$}}}  
\def\B{{\mbox{\boldmath $B$}}}         
\def\noi{\noindent}
\def\nn{\noindent}
\def\sm{${\cal{S}} {\cal{M}}\;$}
\def\nph{${\cal{N}} {\cal{P}}\;$}
\def\sb{$ {\cal{S}}  {\cal{B}}\;$}
\def\ssb{${\cal{S}} {\cal{S}}  {\cal{B}}\;$}
\def\ssbe{{\cal{S}} {\cal{S}}  {\cal{B}}}
\def\cviol{${\cal{C}}\;$}
\def\pviol{${\cal{P}}\;$}
\def\cpviol{${\cal{C}} {\cal{P}}\;$}

\newcommand{\lgg}{\lambda_1\lambda_2}
\newcommand{\lww}{\lambda_3\lambda_4}
\newcommand{\ppin}{ P^+_{12}}
\newcommand{\pmin}{ P^-_{12}}
\newcommand{\ppout}{ P^+_{34}}
\newcommand{\pmout}{ P^-_{34}}
\newcommand{\sinsq}{\sin^2\theta}
\newcommand{\cossq}{\cos^2\theta}
\newcommand{\yt}{y_\theta}
\newcommand{\hppll}{++;00}
\newcommand{\hpmll}{+-;00}
\newcommand{\hpplt}{++;\lambda_30}
\newcommand{\hpmlt}{+-;\lambda_30}
\newcommand{\hpptt}{++;\lambda_3\lambda_4}
\newcommand{\hpmtt}{+-;\lambda_3\lambda_4} 
\newcommand{\dk}{\Delta\kappa}
\newcommand{\klam}{\Delta\kappa \lambda_\gamma }
\newcommand{\kac}{\Delta\kappa^2 }
\newcommand{\lac}{\lambda_\gamma^2 }
\def\gamgamtzz{$\gamma \gamma \ra ZZ \;$}
\def\gamgamtww{$\gamma \gamma \ra W^+ W^-\;$}
\def\gamgamtwwe{\gamma \gamma \ra W^+ W^-}

\setcounter{section}{1}

\setcounter{subsection}{0}
\setcounter{equation}{0}
\def\thesubsection {\thesection.\arabic{subsection}}
\def\theequation{\thesection.\arabic{equation}}
\setcounter{equation}{0}
\def\thequation{\thesection.\arabic{equation}}

\setcounter{section}{0} 
\setcounter{subsection}{0}

\section{Introduction}
During the past few years there has been an intense activity in the
physics potentials of a linear \epemt collider which, in a first stage, 
could be operating at centre-of-mass energies around the 
$t\bar{t}$ threshold to 500~GeV with a later 
upgrade to 1.5-2~TeV\cite{NLCgeneral}. The clean environment of the machine will definitely settle the issue 
of a light Higgs and allow for precision measurements to be conducted. Most prominent among these, 
especially  if no Higgs has been  discovered, is a detailed investigation of the dynamics  of the weak
bosons since they are a window on the mechanism of symmetry breaking through their
Goldstone/longitudinal component\cite{Moijapan}.\\
Another  attraction of the linear collider is that it can be
turned into a \gag collider through Compton scattering of a laser beam on the single pass 
electrons\cite{PhotonCol}.
The resulting \gag collider will thus have a centre-of-mass energy corresponding to about as much as
$80\%$ of the total \epemt energy. One can also  arrange to have an almost monochromatic 
spectrum while maintaining a good luminosity. 
Moreover, the photons are easily polarised. In this type of colliders one of
the most important processes is \ggwwt with a cross section reaching very quickly, after threshold, 
a plateau of about 80~pb. With a contemplated luminosity of about 20~fb$^{-1}$ at 500~GeV this
would mean some millions of $W$ pairs. 
This statistics will thus allow a very precise measurement of
the photonic couplings of the $W$, a subject that has been widely studied
\cite{Parisgg,nousggvv,Yehudai,Choi91,belcou,Hagiwarachoi}. 

The reaction \ggwwt has been studied by various authors and the 
one-loop radiative corrections have already been calculated\cite{ggwwrc}. 
For a more thorough investigation of this most important reaction at the photon collider, 
it is essential to evaluate the 4-fermion final states in \gag processes. 
The full evaluation of these final states is important because they 
are not only reached by 
\ggwwt  but also through contributions that do not proceed {\it via} the 
 doubly resonant $WW$ channel. The latter
could be considered as a potentially important background to \ggwwt and therefore stand in the
way of  a precise investigation of \ggwwt. The same need for the evaluation of all 4-fermion final 
states was  required for LEP2. Consequently a large number of groups undertook the task of
 providing as a
precise as possible calculation of the 4-fermion cross-sections in \epemt\cite{smplep2,evgenlep2}. The aim of this paper 
is to do the
same for \gag initiated 4-fermion processes specialising first in semi-leptonic final states 
$l \nu q \bar{q'}$ for energies currently being discussed for the next linear collider. 
Prior to this paper there 
 appeared only one  evaluation of $\gamma \gamma \ra l \nu q \bar{q'}$ processes 
by Moretti\cite{Morettiggsemilep}. Moreover, the latter has not addressed the important issue of 
the photon polarisations 
and no attempt has been made to compare the results of the 
full 4-fermion calculations including all possible diagrams with some approximation based on the 
$WW$ resonant diagrams. We will therefore give a special attention to the photon 
polarisation and will show that once moderate cuts are applied, a calculation based on \ggwwt 
and which takes  into account the full spin correlations can reproduce, at the per-cent level or even
better, the result of the complete 
set of diagrams after allowing for a smearing factor. We will also consider  different schemes for 
the implementation of the $W$ width and whenever appropriate we will compare our results with those 
of Moretti\cite{Morettiggsemilep}\footnote{Total cross sections for purely QED 4-fermion production
through \gag have been considered for quite some time\cite{qedgg4f}. Leptonic final states 
of the kind $\gamma \gamma \ra e^- \bar{\nu}_e \mu^+ \nu_{\mu}$ have first been evaluated by 
Couture\cite{Couturegg4f}. The (unpolarised) results for the other leptonic channels have very recently been 
presented in \cite{Morettigg4f}. }. \\

\section{On-Shell $W$ pair production and decay with full spin correlations and smearing} 
\setcounter{secnumdepth}{2} 
\setcounter{equation}{0}
\def\thesubsection{\thesection.\arabic{subsection}} 
\def\theequation{\thesection.\arabic{equation}} 
\subsection{Tree-level helicity amplitudes for $\gam \ra 
W^+W^-$ in the \sm} 
To understand the characteristics of the $W^+W^-$ cross-section it is best to 
give all the helicity amplitudes which 
contain a maximum of information on the reaction.

It is important to specify our conventions. We work in the centre of mass 
of the incoming photons and refrain from making explicit the azimuthal dependence 
of the initial state. The total energy of the \gag system is $\sqrt{s}$. 
We take the photon with helicity $\la_1$ ($\la_2$) to be in the $+z$ ($-z$) direction 
and the outgoing 
$W^-$ ($W^+$) with helicity $\la_-$ ($\la_+$) and 4-momentum $p_-$  ($p_+$):
\beqn
p_\mp^\mu=\frac{\sqrt{s}}{2}(1, \pm \beta \sin \theta,0, \pm\beta \cos \theta)
\;\;\; ; \;
\;\; \beta=\sqrt{1-4/\gamma} \; ; \; \gamma=s/\mww.
\eeqn 

The polarisations for the helicity basis are defined as 
\beqn
\epsilon_1^\mu(\la_1)=\frac{1}{\sqrt{2}} (0,-\la_1,-i,0) \;\;\;\;\;\;\;\;\;& &
\epsilon_2^\mu(\la_2)=\frac{1}{\sqrt{2}} (0,\la_2,-i,0)    \;\;\;\; \la_{1,2}=\pm  \\
\epsilon_-^\mu(\la_-)^*=\frac{1}{\sqrt{2}} 
(0,-\la_- \cos \theta,i,\la_- \sin \theta)& &
\epsilon_+^\mu(\la_+)^*=\frac{1}{\sqrt{2}} 
(0,\la_+ \cos \theta,i,-\la_+ \sin \theta) \;\;\;\; \la_{\pm}=\pm \nonumber \\ 
\epsilon_-^\mu(0)^*=\frac{\sqrt{s}}{2M_W} (\beta, \sin \theta,0,\cos \theta)
\;\;\;\;\;\;\;\;\;& &
\epsilon_+^\mu(0)^*=\frac{\sqrt{s}}{2M_W} (\beta, -\sin \theta,0,-\cos \theta) \;\;\;\;
\la_{\pm}=0.\nonumber
\eeqn
We obtain for the tree-level \sm helicity amplitudes
\footnote{These formulae had already been derived  by us\cite{Parisgg,nousggvv} 
albeit with a different convention for 
the polarisation vectors. They are also consistent with those 
given by Yehudai\cite{Yehudai} and more recently by 
Choi and Hagiwara\cite{Hagiwarachoi} after allowing for the different conventions for the 
polarisation vectors.}:
\beq
{\cal M}_{\la_1 \la_2; \la_- \la_+} = \frac{4 \pi \alpha}
{1-\beta^2 \cos^2 \theta} \;\;{\cal N}_{\la_1 \la_2; \la_- \la_+},
\label{eq:mn}
\eeq
\noindent
where
\beqn
{\cal N}_{\la_1 \la_2; 0 0} &= &\;-\;\frac{1}{\gamma}
\left\{ 
-4 (1+\la_1 \la_2) + (1-\la_1 \la_2) (4+\gamma) \sin^2 \theta  \right\},
 \\
{\cal N}_{\la_1 \la_2; \la_- 0} &=& \sqrt{\frac{8}{ \gamma}} \;\; (\la_1-\la_2) 
         (1+\la_1 \la_- \cos \theta) \sin \theta,  \;\;\;\; \la_-=\pm \nonumber \\
{\cal N}_{\la_1 \la_2; 0,\la_+ } &=&\;-\;\sqrt{\frac{8}{ \gamma}} \;\; (\la_1-\la_2) 
         (1-\la_1 \la_+ \cos \theta) \sin \theta,
 \;\;\;\; \la_+=\pm \nonumber  \\
{\cal N}_{\la_1 \la_2; \la_- \la_+}&=& \beta (\la_1+\la_2) (\la_-+\la_+) + 
\frac{1}{2 \gamma}\left\{
-8 \la_1 \la_2 (1+\la_- \la_+) +\gamma (1+\la_1 \la_2 \la_- \la_+) 
(3+\la_1 \la_2) \right.  \nonumber \\
&+& \left.  2 \gamma (\la_1-\la_2)(\la_--\la_+)\cos\theta
- 4 (1-\la_1 \la_2)(1+\la_- \la_+) \cos^2\theta \right. \nonumber \\
&+& \left. \gamma (1-\la_1 \la_2)(1-\la_- \la_+) \cos^2\theta \right\}
\;\;\;\; \;\;\;\; \;\;\;\; \la_\pm=\pm.
\label{eq:lesbons} 
\eeqn

With the conventions for the polarisation vectors, the fermionic tensors that describe the decay of the 
$W$'s are defined 
as in \cite{Fernandeeww}. In particular one expresses 
everything with respect to the $W^-$ where the arguments of the $D$ functions 
refer to the angles of  the particle 
({\em i.e.} the electron, not the anti-neutrino),
in the rest-frame of the $W^-$, taking as a reference axis 
the direction of flight of the $W^-$ (see~\cite{Fernandeeww}). The $D$-functions to use are therefore
$D^{W^-}_{\la,\la'}(\theta^*, \phi^*)\equiv D_{\la,\la'}$, satisfying 
$D_{\la,\la'}=D_{\la',\la}^*$  and:
\beqn
\label{dfunctions}
D_{+,-}=\frac{1}{2} (1-\cos^2 \theta^*) e^{2i\phi^*},& &
D_{\pm,0}=-\frac{1}{\sqrt{2}} (1 \mp \cos \theta^*)\sin \theta^* e^{\pm i \phi^*}, \\ \nonumber
D_{\pm,\pm}=\frac{1}{2} (1  \mp \cos \theta^*)^2,& &D_{0,0}=\sin^2 \theta^*.
\eeqn
The angle $\theta^* \equiv \theta^*_e$ is directly related to the energy of the electron (measured in 
the laboratory frame):
\beqn
\label{thetastar}
\cos\theta^*_e=\frac{1}{\beta} \left( \frac{4 E_e}{\sqrt{s}} -1 \right).
\eeqn

There are a few remarks 
one can make about the structure of the 
\ggwwt helicity amplitudes. First of all, because of the t-channel 
spin-1 exchange, the $W$'s are produced predominantly 
in the forward/backward direction. An effect that is more pronounced as the 
energy increases. 
The production of longitudinal $W$'s occurs predominantly when the photons are in a 
$J_Z=2$ configuration. Otherwise the amplitudes for longitudinal 
$W$'s are at least a factor $M_W/\sqrt{s}$ smaller than the amplitudes for  
transverse $W$'s. Another very interesting property is that for like-sign 
photon helicities only like-sign $W$ helicities are produced. Moreover,  
as the energy increases this occurs with the photons transferring their 
helicities to the $W$'s. Therefore for like-sign photon helicities, the cross section is 
dominated by 
$\sigma_{+,+;+,+}$ and $\sigma_{-,-;-,-}$.

\subsection{Five-fold differential cross section}

The five-fold differential cross section can now be easily obtained. In the narrow width
approximation and for definite photon helicities as defined above the five-fold differential 
cross section $\gamma \gamma \ra f_1 \bar{f_2}\; f_3 \bar{f_4}$ writes
\beqn
\label{fullspincorr}
& & 
\mbox{}\frac{ {\rm d}\sigma(\gamma(\lambda_1)\gamma(\lambda_2) \ra W^+W^-\ra f_1 \bar f_2 f_3 \bar f_4)}
     { {\rm d}\cos \theta \;\;{\rm d}\cos \theta_-^{*} \;\;{\rm d}\phi_-^{*}\;\;
      {\rm d}\cos \theta_+^{*} \;\;{\rm d}\phi_+^{*} }=Br^{f_1 \bar f_2}_W Br^{f_3 \bar f_4}_W 
\frac{\beta}{32\pi s}
\left( \frac{3}{8 \pi}\right)^2
\hfill \nonumber \\
\lefteqn{ \sum_{\lambda_- \lambda_+ \lambda'_- \lambda'_+}
 {\cal M}_{\lambda_1,\lambda_2; \lambda_-\lambda_+} (s,\cos \theta)\;  
{\cal M}_{\lambda_1,\lambda_2; \lambda'_-\lambda'_+}^{*} (s,\cos \theta)
\; D_{\lambda_- \lambda'_-} (\theta_-^{*} ,\phi_-^{*}) \; D_{\lambda_+ \lambda'_+} (\pi-\theta_+^{*},
\phi_+^{*}+\pi)}
\nonumber \\
\lefteqn{ 
\equiv 
\frac{{\rm d}\sigma(\gamma(\lambda_1)\gamma(\lambda_2) \ra W^+W^-)}{{\rm d}\cos \theta}
\left( \frac{3}{8 \pi}\right)^2 Br^{f_1 \bar f_2}_W Br^{f_3 \bar f_4}_W
} \nonumber \\
\lefteqn{ 
\sum_{\lambda_- \lambda_+ \lambda'_- \lambda'_+}
\rho_{\lambda_- \lambda_+ \lambda'_- \lambda'_+}^{\lambda_1,\lambda_2}\;
D_{\lambda_- \lambda'_-} (\theta_-^{*} ,\phi_-^{*}) \;
D_{\lambda_+ \lambda'_+} (\pi-\theta_+^{*}, \phi_+^{*}+\pi) 
}
\nonumber \\
&{\rm with}& \hspace*{.6cm}\rho_{\lambda_- \lambda_+ \lambda'_- \lambda'_+}^{\lambda_1,\lambda_2}(s, \cos\theta)=
\frac{ {\cal M}_{\lambda_1,\lambda_2; \lambda_-\lambda_+} (s,\cos \theta)\;  
{\cal M}_{\lambda_1,\lambda_2; \lambda'_-\lambda'_+}^{*} (s,\cos \theta)} 
{ \sum_{\lambda_- \lambda_+} |{\cal M}_{\lambda_1,\lambda_2; \lambda_-\lambda_+}(s,\cos \theta)|^2},
\eeqn
\noindent where $\theta$ is the scattering angle of the $W^-$ and $\rho$ is 
the density matrix.  
Note that one separates the decay and the production parts. 
To be able to reconstruct the direction of the charged $W$ and to have least 
ambiguity in reconstructing the $WW$ process, 
the best channel is the semi-leptonic channel 
$W^\pm\ra l \nu_l, W^\mp\ra jj'$ with $l=e,\mu$.

This very compact formula makes it possible to impose cuts on the escape angles and energies of the
fermions. However,
because of the nature of this narrow width approximation which is actually a zero-width
approximation, the invariant mass of each fermion pair is equal to 
the W-mass. 

 To improve 
the above approximation and to make it possible to take into account cuts on the invariant mass 
of the decay products of the $W^\pm$ one can introduce a smearing over the invariant masses, 
$s_\pm$, of the $W^\pm$. Customarily one 
has \cite{Japaneewwwidth,wwsmlep2}, as with the  \eewwt cross section,  convoluted over 
the doubly {\em off-shell} 
\ggwwt cross section:
\beqn
\label{offshellcross}
\tilde{\sigma}_{\Gamma_W}=\int_{0}^{s} ds_- \; \rho_W(s_-) \int_{0}^{(\sqrt{s}-\sqrt{s_-})^2} 
 ds_+ \; \rho_W(s_+)\; \sigma_{\ggww}(s_+,s_-),
\eeqn
with 
\beq
\rho_W(x)=\frac{1}{\pi} \frac{M_W \Gamma_W}{(x-M_W^2)^2+M_W^2\Gamma_W^2}.
\eeq
This prescription is  not gauge invariant. $\sigma_{\ggww}(s_+,s_-)$ is derived from
an off-shell amplitude which is not an element of 
the $S$-matrix.
Although $\sigma_{\ggww}(s_+,s_-)$ refers pictorially to doubly ``resonant" diagrams, it 
contains pieces that are only singly resonant, beside other parts that are not resonant at all. To
wit, one can expand $\sigma_{\ggww}(s_+,s_-)$ as
\beqn
\label{offshellwwamp}
 \sigma_{\ggww}(s_+,s_-)&=&\sigma_{\ggww}(s_+=M_W^2,s_-=M_W^2)  \nonumber \\
&+&(s_- - M_W^2) S(s_+) + 
(s_+ - M_W^2) S(s_-) + N(s_+,s_-).
\eeqn
The first part, $\sigma_{\ggww}(s_+=M_W^2,s_-=M_W^2)$ is the genuine on-shell \ggwwt cross section
which is gauge invariant and that we have used to derive the density matrix. 
However the functions $S$ and $N$ are not gauge invariant and must be
combined with the complete set of the 4-fermion diagrams. A problem that we will address in the next
section. 
Therefore to improve on the approximation for the density matrix,
we will simply convolute the on-shell fully correlated cross section Eq.~\ref{fullspincorr}
$\sigma_{\ggww}(s_+=M_W^2,s_-=M_W^2)\equiv \sigma_{\ggww}(s)$, to obtain 
\beqn
\sigma_{\Gamma_W}&=&R_\Gamma \;\sigma_{\ggww}(s) \nonumber \\ 
&=&  \sigma_{\ggww  \ra f_1 \bar f_2 f_3 \bar f_4}(s) 
\int_0^s ds_- \; \rho_W(s_-)\; \int_0^{(\sqrt{s}-\sqrt{s_-})^2}\; F_{cut}(s_+,s_-)\;
 ds_+ \; \rho_W(s_+), \nonumber \\
\eeqn
where we have  introduced a function $F_{cut}$ which takes into account the cuts that we may impose 
on the invariant masses, $s_\pm$. $R_\Gamma$ is the reduction factor introduced by the width and the
invariant mass cuts. 
The above formula can also be applied when other cuts on the decay products 
are imposed, as long as these cuts do not implicitly or explicitly involve $s_\pm$. In this 
case the factorisation holds with $\sigma_{\ggww}(s)$ calculated through the fully correlated 
amplitudes.

For invariant mass cuts such that, 
\beq
\label{invmasscutdef}
|\sqrt{s_\pm} -M_W|\leq \Delta,
\eeq
one has 
\beqn
R_\Gamma&\simeq&\frac{1}{\pi^2} \left\{ Atan\left( \frac{\Delta (2M_W+\Delta)}{\Gamma_W M_W}\right) 
\;+\;
 Atan\left( \frac{\Delta (2M_W-\Delta)}{\Gamma_W M_W}\right) 
\right\}^2   \nonumber \\ 
& \sim& 
\left( \frac{2}{\pi} Atan\left( \frac{2\Delta}{\Gamma_W} \right) \right)^2 \;\;\;\; 
(\Delta \ll 2M_W).
\eeqn

If the invariant mass cuts on $s_\pm$ are different, $|\sqrt{s_\pm} -M_W|\leq \Delta_\pm$, the
reduction factor is the product of the square root of the above formulae with $\Delta \ra \Delta_+$
for one of the factor and  $\Delta \ra \Delta_-$ in the second. However if one imposes an invariant
mass cut on only one of the $W$, a very good approximation for the reduction factor is

\beqn
R_\Gamma&\simeq&\frac{1}{\pi^2} \left\{ Atan\left( \frac{\Delta (2M_W+\Delta)}{\Gamma_W M_W}\right) 
\;+\;
 Atan\left( \frac{\Delta (2M_W-\Delta)}{\Gamma_W M_W}\right) 
\right\} \times \nonumber \\ 
&& \left\{ Atan\left( \frac{M_W}{\Gamma_W} \right) + 
Atan\left( \frac{(\sqrt{s}-2M_W)\sqrt{s}}{\Gamma_W M_W } \right) \right\}
\eeqn

In the case where no cut on  the invariant masses 
is imposed, to take the width effect into account  one can apply the approximate formula 
\beqn
R_\Gamma \simeq 1\;-\; \frac{1}{\pi} \frac{\Gamma_W}{M_W} \left( 2+ \frac{M_W^2}{s-M_W^2} 
+\frac{M_W^2}{s-2M_W\sqrt{s}} \right).
\eeqn

\section{Complete calculation of four-fermion production}
\begin{figure*}[htbp]
\begin{center}
\mbox{\epsfxsize=15cm\epsfysize=7cm\epsffile{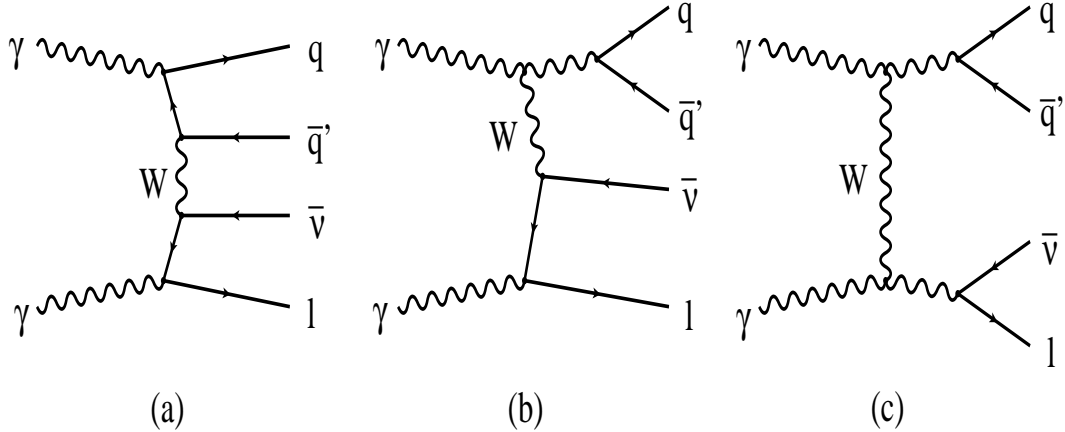}}
 \vspace*{-0.8cm}
\caption[.]{ {\em Classes of Feynman diagrams that contribute to the semi-leptonic
four-fermion production in $\gamma \gamma$. In the narrow width approximation 
only diagrams of type (c) is considered. Type (b) may be associated to single $W^+$ production while 
type (a) are non-resonating.} \label{figgg4f}}
\end{center}
\end{figure*}

The tree-level calculation of $\gamma \gamma \ra e^- \bar{\nu}_e u \bar{d}$ involves 
the evaluation of 21 diagrams in the unitary gauge (see. Fig.~\ref{figgg4f}), while the narrow width approximation 
is based on a mere 3 diagrams. 
As we have seen, the narrow width approximation can be improved by taking into account 
finite width effects even when cuts on the invariant mass of the decay products of the 
$W$'s are envisaged. Since this approximation leads to very compact analytical formulae 
it is important to quantify how good this approximation can be compared to the full 
calculation of the 4-fermion final state. \\

Beside those going through $WW$, 
some of the latter 
4-fermion final states may be associated to single $W$ production of the sort
$\gamma \gamma \ra e^- \bar{\nu}_e W^+$ with the subsequent decay of the $W^+$ 
( see Fig.~\ref{figgg4f}, diagram (b))
. This type of
contribution is 
 important for very forward electrons with a typical escape angle for 
the electron of order $m_e/\sqrt{s}$. For these very forward events  it should be possible 
to revert to an approximation based on a structure function approach 
(density of electrons in the photon). We will not be concerned with these types of events 
since our motivation is to fully 
reconstruct the $WW$ events which requires 
the observation of the charged fermion as well as the jets in the central region.  
This is especially important if one
wants to investigate New Physics effects affecting the $W$ sector through \ggwwt. 
These require one
to be able to fit angular distributions of the fermions through, for example,  a maximum likelihood
method\cite{nousggwwml}. \\
As for events with  very forward jets, another complication arises if very 
forward jet-tagging is not possible. In this situation one needs to consider 
other contributions beyond the 21 diagrams leading to  $\gamma \gamma \ra e^- \bar{\nu}_e u
\bar{d}$. These additional contributions 
are due to  the non-perturbative hadron component of the photon and thus 
need a special treatment.
We therefore concentrate on the 
$l \bar{\nu}_l u \bar{d}$ (and its conjugate) signature with observable charged particles. We have 
therefore imposed a cut on the charged fermions such that:
\beq
|\cos \theta_{l,j}|<0.98\;\;\;\;\;\;\;\;\;\;\;\;\;\;\;\;\;\;
\cos <l,j><0.9 
\eeq
Moreover we also imposed a cut on the energies of the charged fermions:
\beq
E_f > 0.0125 \sqrt{s}.
\eeq

We take $\alpha=\alpha(0)=1/137$ for the $WW\gamma$ vertex as 
we are dealing with an on-shell photon.
The result of the full one-loop corrections to 
\ggwwt \cite{ggwwrc} indicates also that this is the effective coupling constant for \ggwwt.
We also take $M_W=80.22$~GeV
and the width of the 
$W$, is $\Gamma_W=\Gamma_W(M_W^2)=2.08$~GeV. 
The partial width  of the $W$ into jets and $l \bar{\nu_l}$ is 
{\em calculated} by taking at the $W$ vertex the effective couplings 
$\alpha(M_W^2)=1/128$ and $\sin^2\theta_W=0.23$.   

In order to quantify the effect of the ``non doubly resonant" contributions and how well the 
approximation based on the density matrix approach fares, 
we have considered different cuts on the invariant mass $\Delta$ (see~\ref{invmasscutdef}) of the jets 
($\Delta_{jj}$) and the 
leptonic pair ($\Delta_{l\bar{\nu}_l}$). Especially because for a realistic $\gamma \gamma$ set-up where the
photon energy is not known in an event by event basis, we have also relaxed the cut on the leptonic
invariant mass. Two values on the invariant mass of the pairs were considered. A stringent cut
$\Delta=5$~GeV$\sim 2\Gamma_W$
and a cut that is more realistic when dealing with a jet pair: $\Delta_{jj}=18$~GeV. 

There is another issue that needs to be discussed before we present our results. This concerns the 
implementation of the width in the $W$ propagators. This important issue has received quite a
lot of attention recently in connection with 4-fermion production in 
\epemt\cite{wwsmlep2,widthissue,FLscheme}. 
The introduction of the
width is necessary if one is to integrate over all virtualities of the $W$'s. 
The naive way of 
providing the $W$ propagators either with a running or fixed width does not lead to a
gauge-invariant 4-fermion matrix element, since only one subset (the one corresponding to 
the ``resonant" diagrams) of the whole 4-fermion diagrams 
has received a ``correction". Prior to the introduction of the width, the \underline{complete} 
4-fermion amplitude is gauge invariant, while the resonant diagrams are gauge invariant by themselves only 
when the $W$ are both on mass-shell as calculated in the previous section.  
To remedy this situation, introduced by ``correcting" the two-point function, it has been suggested to 
implement radiative corrections to the three-point as well as to the four-point 
vertices\cite{Baurwidth,widthissue,FLscheme}. 
This is quite
cumbersome to implement on the 4-fermion final state and at the Monte-Carlo level, moreover these
additional corrections need to be carried at all orders in perturbation theory while up to now this 
approach has been implemented only at the one-loop level. 
Different other schemes or rather simplified tricks 
have been suggested and implemented in most of the $e^+e^- \ra 4f$ existing programs (for a review
see~\cite{wwsmlep2,widthissue,FLscheme}), but on a 
formal level none is satisfactory.
On a practical level, it is found\cite{wwsmlep2} , for \eewwt, that 
\underline{after} implementing 
cuts (in our case tagging the jets and the electron/positron) all schemes agree quite well. 
To quantify
the sensitivity to the $W$ width implementation we have implemented  a fixed width, a
running width as well as what is called a ``fudge-factor"\cite{fudgefactor}. The fudge factor amounts to first
generating all the 4-fermion helicity amplitudes with zero width, that is the $W$ propagators 
have the simple form $1/(k^2-M_W^2)$ and thus gauge invariance is guaranteed. One then applies 
 an overall
fudge-factor $(k^2-M_W^2)/(k^2-M_W^2+i \Gamma_W M_W)$\footnote{One may also apply a running width
here.} to the whole amplitude.   Since we consider $\gamma \gamma$ energies where the doubly-resonant diagrams give a
large contribution this fudge factor should be a good approximation. 
Related to this issue we have 
also investigated the effect of picking up among the 21 4-fermion diagrams, those three that
correspond to the double $WW$ exchange, where the $W$ may be off-shell. The amplitudes thus 
generated lead to the ``off-shell cross sections" defined in Eq.~\ref{offshellcross}. As we argued
above, unless we restrict 
$s_\pm$ to be very close to $M_W^2$, we expect gross deviations from the result of the evaluation of
the full set of diagrams. This  will be clearly substantiated in our results.

The helicity amplitudes based on the full set of contributions were reproduced through the automatic 
program {\em Madgraph}\cite{madgraph} and checked with {\em GRACE}\cite{grace}. The numerical phase space integration
was done  with the help of {\em VEGAS}\cite{vegas}. As a first comparison,
we have tuned our set of input parameters and cuts to those chosen by Moretti\cite{Morettiggsemilep}
 and have reproduced
his results for the unpolarised cross sections.

\subsection{Results and Comparison}

\begin{table}[htbp]\centering
\begin{center}
\begin{tabular}{|c|c||c||c|c|c||c|} \hline
\multicolumn{7}{|c|}{{\bf $\sqrt{s}=400$~GeV}}\\ \hline
$\lambda_1\;\lambda_2$& Inv. Mass. & Narrow Width & All diag.& All diag. & All diag.&
``Resonant" \\ 
&Cuts&Improved&$\Gamma_W(M_W^2)$&$\Gamma_W(s)$&Fudge&Subset\\
\hline \hline
&&&&&&\\
$+\;+$&  None &  2288&   2310 &  2312 &  2309&   2354\\
$+\;-$  & None&   1893	 & 	1926  &   1927 & 	1923  &   1975\\
$-\;+$  & None  &   1890 & 	1927  &   1927 & 	1926 &   1975\\
$-\;-$  &  None &   2186  & 	2184  &   2183 & 	2183  &   2252\\
&&&&&&\\
$+\;+$ & $\Delta_{jj},\Delta_{l\nu} < 5$~GeV&   1759 & 1762  & 1764 & 	1761  &   1761\\
$+\;-$  & $\Delta_{jj},\Delta_{l\nu} < 5$~GeV  &   1455  & 	1458  &  
1458 & 1456 &   1456\\
$-\;+$  & $\Delta_{jj},\Delta_{l\nu} < 5$~GeV  &  1454 & 1457  &  1458 & 1457   &  1456\\
$-\;-$  & $\Delta_{jj},\Delta_{l\nu} < 5$~GeV  &   1681  & 	1683  &   1681  & 1682  &   
1682\\
&&&&&&\\
$+\;+$ &$\Delta_{jj},\Delta_{l\nu} < 18$~GeV  &   2158  & 	2167   &  2169  & 	2165  &  
2166\\
$+\;-$  &$\Delta_{jj},\Delta_{l\nu} < 18$~GeV &   1785  & 	1793 &  
1794  & 
1791  &   1793\\
$-\;+$ & $\Delta_{jj},\Delta_{l\nu} < 18$~GeV  &     1783  & 	1795   &  1794  & 
1794  &   1793\\
$-\;-$  &$\Delta_{jj},\Delta_{l\nu} < 18$~GeV &  2061 &	2063 &  2061&	2061&  
2068\\
&&&&&&\\
$+\;+$ &$\Delta_{jj}< 5$~GeV &  2006&	2026  & 2027&	2024 &  2035\\
$+\;-$    &$\Delta_{jj}< 5$~GeV&  1660&	1682  & 1682&	1679 & 
1695\\
$-\;+$ &$\Delta_{jj}< 5$~GeV  &  1658&	1683  & 1683&	1683  & 1696\\
$-\;-$   &$\Delta_{jj}< 5$~GeV  &  1917&	1921 &  1920&	1921 &  1948\\
&&&&&&\\
$+\;+$&$\Delta_{jj}< 18$~GeV  &  2222 &	2245 &  2247 & 2243 &  2257\\
 $+\;-$  &$\Delta_{jj}< 18$~GeV  & 1838&	1865   & 1866 &
1863  &   1881\\
$-\;+$  & $\Delta_{jj}< 18$~GeV &   1836 &	1866   & 1866&	1865 &  1881\\
$-\;-$ &$\Delta_{jj}< 18$~GeV   &  2123 &	2127  &  2125 &	2125  & 
2159\\
&&&&&&\\
\hline \hline  
\end{tabular}
\end{center}
\caption[.]{{\em Comparison between the different approximations for the calculation of 
$\gamma \gamma \ra e^- \bar{\nu}_e u \bar{d}$ for a centre-of-mass energy of 400~GeV. 
$\lambda_{1,2}$ refer to the helicities of the two 
photons. Cuts other than those on the invariant mass are given in the text. ``Narrow width
Improved" refers to the approximation based on the density matrix and takes into account smearing. 
The next column is the result based on the complete set of the 4-fermion diagrams with a fixed
width. ``All diag." $\Gamma_W(s)$ is with a running width, while ``Fudge" implements a fudge factor 
as explained in the text. The last column gives the result of keeping only the $WW$ 4-fermion
diagrams, {\it i.e.} based on the off-shell \ggwwt amplitudes (where  a fixed with is used). All cross
sections are in fb.}}
\label{tableone}
\end{table}
\normalsize

\begin{table}[htbp]\centering
\begin{center}
\begin{tabular}{|c|c|c|c|c|c|c|} \hline
\multicolumn{7}{|c|}{{\bf $\sqrt{s}=800$~GeV}}\\ \hline
$\lambda_1\;\lambda_2$& Inv. Mass. & Narrow Width & All diag.& All diag. & All diag.&
``Resonant" \\ 
&Cuts&Improved&$\Gamma_W(M_W^2)$&$\Gamma_W(s)$&Fudge&Subset\\
\hline \hline
&&&&&&\\
$+\;+$&  None & 1274  &  1290  &  1290  &  1287  &  1399 \\
$+\;-$  & None&  1010  &	1034   & 1032 &	1032  &  1119\\
$-\;+$  & None&   1009  &	1037 &   1035& 	1036 &  1121\\
$-\;-$  & None& 1232 & 	1260  &   1259&	1259 &  1335\\
&&&&&&\\
$+\;+$ & $\Delta_{jj},\Delta_{l\nu} < 5$~GeV& 979 &	983  & 982&
982 &  983\\
$+\;-$ & $\Delta_{jj},\Delta_{l\nu} < 5$~GeV&  777 &	776 &  775&
776  & 776\\
$-\;+$ & $\Delta_{jj},\Delta_{l\nu} < 5$~GeV& 776 &	777 &  777&
777 &  778\\
$-\;-$ & $\Delta_{jj},\Delta_{l\nu} < 5$~GeV& 947 &	950 &  947&	948 & 
950\\
&&&&&&\\
$+\;+$ & $\Delta_{jj},\Delta_{l\nu} < 18$~GeV&  1201 &	1207 &  1205&
1203 &  1211\\
$+\;-$ & $\Delta_{jj},\Delta_{l\nu} < 18$~GeV&952& 953 & 951&
952&   956\\
$-\;+$ & $\Delta_{jj},\Delta_{l\nu} < 18$~GeV&951& 954&   954&	954&  958\\
$-\;-$ & $\Delta_{jj},\Delta_{l\nu} < 18$~GeV&1162 & 1164&   1162&
1163 &  1167\\
&&&&&&\\
$+\;+$ & $\Delta_{jj} < 5$~GeV&  1117 &1128&   1127&1126&   1164\\
$+\;-$ & $\Delta_{jj} < 5$~GeV&  886 &	899&  898&899 & 
926\\
$-\;+$ & $\Delta_{jj} < 5$~GeV&  885 &	902 &  901&901 & 
928\\
$-\;-$ & $\Delta_{jj} < 5$~GeV& 1080 &	1097 &  1095&1096&   1123\\
&&&&&&\\
$+\;+$ & $\Delta_{jj} < 18$~GeV& 1237 &	1250 &  1249&	1246 &  1294\\
$+\;-$ & $\Delta_{jj} < 18$~GeV&   981& 	996&   994&	995 &  1029\\
$-\;+$ & $\Delta_{jj} < 18$~GeV&  980 &	999&   998&	998 &  1031\\
$-\;-$ & $\Delta_{jj} < 18$~GeV&  1197 &	1214 &  1213&	1213& 
1247\\
&&&&&&\\
\hline \hline 
\end{tabular}
\end{center}
\caption[.]{{\em As in Table~1 but for $\sqrt{s}$=800~GeV.}}
\label{tabletwo}
\end{table}
\normalsize

\begin{table}[htbp]\centering
\begin{center}
\begin{tabular}{|c|c|c|c|c|c|c|} \hline
\multicolumn{7}{|c|}{{\bf $\sqrt{s}=1600$~GeV}}\\ \hline
$\lambda_1\;\lambda_2$& Inv. Mass. & Narrow Width & All diag.& All diag. & All diag.&
``Resonant" \\ 
&Cuts&Improved&$\Gamma_W(M_W^2)$&$\Gamma_W(s)$&Fudge&Subset\\
\hline \hline
&&&&&&\\
$+\;+$&  None &   377  & 389 &  389  & 389&   456\\
$+\;-$&  None &   320&	335 &  335&	335  & 388\\
$-\;+$&  None &  320&	336 &  336 &	336 & 391\\
$-\;-$&  None &  427&	447 &  447 &	447   &490\\
&&&&&&\\
$+\;+$ & $\Delta_{jj},\Delta_{l\nu} < 5$~GeV&  290&	291 &  291&	291  & 291\\
$+\;-$ & $\Delta_{jj},\Delta_{l\nu} < 5$~GeV&   246&	246 &  246&	246  &
246\\
$-\;+$ & $\Delta_{jj},\Delta_{l\nu} < 5$~GeV&   246&	246  & 247&	246  &
247\\
$-\;-$ & $\Delta_{jj},\Delta_{l\nu} < 5$~GeV&  328&	329 &  329&
330  & 329\\
&&&&&&\\
$+\;+$ & $\Delta_{jj},\Delta_{l\nu} < 18$~GeV&    356&	357 &  357&	357  &
358\\
$+\;-$ & $\Delta_{jj},\Delta_{l\nu} < 18$~GeV&    302& 302 &  302&	302 & 302\\
$-\;+$ & $\Delta_{jj},\Delta_{l\nu} < 18$~GeV&   302&	303 &  303&	303  &
304\\
$-\;-$ & $\Delta_{jj},\Delta_{l\nu} < 18$~GeV&   403&	404 &  404&
405  &   405\\
&&&&&&\\
$+\;+$ & $\Delta_{jj} < 5$~GeV&    331&337 &  336&	337  &  355\\
$+\;-$ & $\Delta_{jj} < 5$~GeV&   281&	288 &  288&	289 &  302\\
$-\;+$ & $\Delta_{jj} < 5$~GeV&    281&289 &  289&	289 &  303\\
$-\;-$ & $\Delta_{jj} < 5$~GeV&    375&385 &  385&	385 &  398\\
&&&&&&\\
$+\;+$ & $\Delta_{jj} < 18$~GeV&     366&373 &  373&	373 &  393\\
$+\;-$ & $\Delta_{jj} <18$~GeV&    311&319 &  319&	320 &  336\\
$-\;+$ & $\Delta_{jj} < 18$~GeV&    311&320 &  320&	320 &  336\\
$-\;-$ & $\Delta_{jj} < 18$~GeV&   415&426&  426&	427 &  442\\
&&&&&&\\
\hline \hline 
\end{tabular}
\end{center}
\caption[.]{{\em As in Table~1 but for $\sqrt{s}$=1600~GeV}}
\label{tablethree}
\end{table}

\normalsize \normalsize 

\normalsize

\baselineskip=18pt

Our results including the different implementations and approximations are presented in 
Tables~\ref{tableone}~--~\ref{tablethree}. Before commenting on the 
comparisons, let us 
stress that for any approximation or implementation of the width,  
even after different cuts are imposed the two $J_Z=2$ configurations 
must be equal, {\it i.e.},  one must have
$\sigma_{+-}=\sigma_{-+}$  which is a direct consequence 
of the fact that the cuts we used are the same in the ``forward'' and 
``backward'' directions and that the initial particles are identical (Bose symmetry).
This is well rendered by our calculations
as evidenced from the Tables. 
The very slight differences (below the per-mil) that the tables show for these
combinations of photon helicities are a measure of the precision of our 
MC (Monte-Carlo) integration. We have prefered 
to keep both the $\sigma_{+-}$ and $\sigma_{-+}$ entries in our tables 
so that their  
relative precision be used as a benchmark for a comparison between the different 
approximations. We have considered three typical $\gamma \gamma$ centre-of-mass energies: 400,800
and 
1600~GeV corresponding to  \epemt centre-of-mass energies of 500, 1000 and 2000~GeV.

\begin{figure*}[hbtp]
\vspace*{-1.5cm}
\begin{center}
\mbox{\epsfxsize=15cm\epsfysize=15cm\epsffile{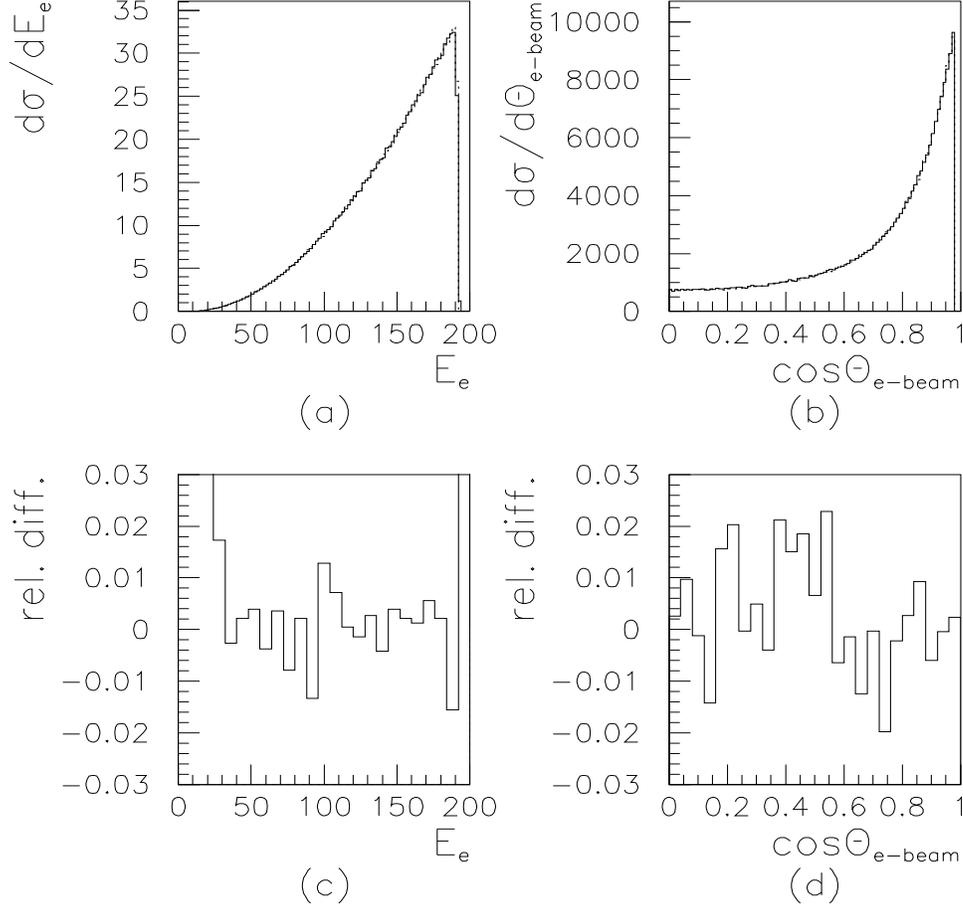}}
 \vspace*{-0.8cm}
\caption{ {\em Comparison between the distributions obtained with  the improved 
narrow width approximation and with the computation based on the complete set of 
diagrams for a centre-of-mass energy of 400GeV with both photons being left-handed 
($\sigma_{--}$). A cut of 18GeV 
on the 
invariant mass of the jets has been imposed. All other cuts are as specified in the text. 
 Shown are the distribution in the energy of the 
electron, $E_e$ (a), and the cosine of the angle of the electron with the beam 
$\cos \Theta_{e-{\rm beam}}$ (b) where the dotted line are for he approximation. 
The differences are hardly noticeable on these distribution. In order to better show the 
discrepancies, we plot in Figs.~(c) and (d) the relative difference between the two 
distributions}
$({\rm Exact} - {\rm Approx.} )/{\rm Exact}$.
\label{figdis400} }
\end{center}
\end{figure*}

\begin{figure*}[hbtp]
\vspace*{-1.5cm}
\begin{center}
\mbox{\epsfxsize=15cm\epsfysize=15.cm\epsffile{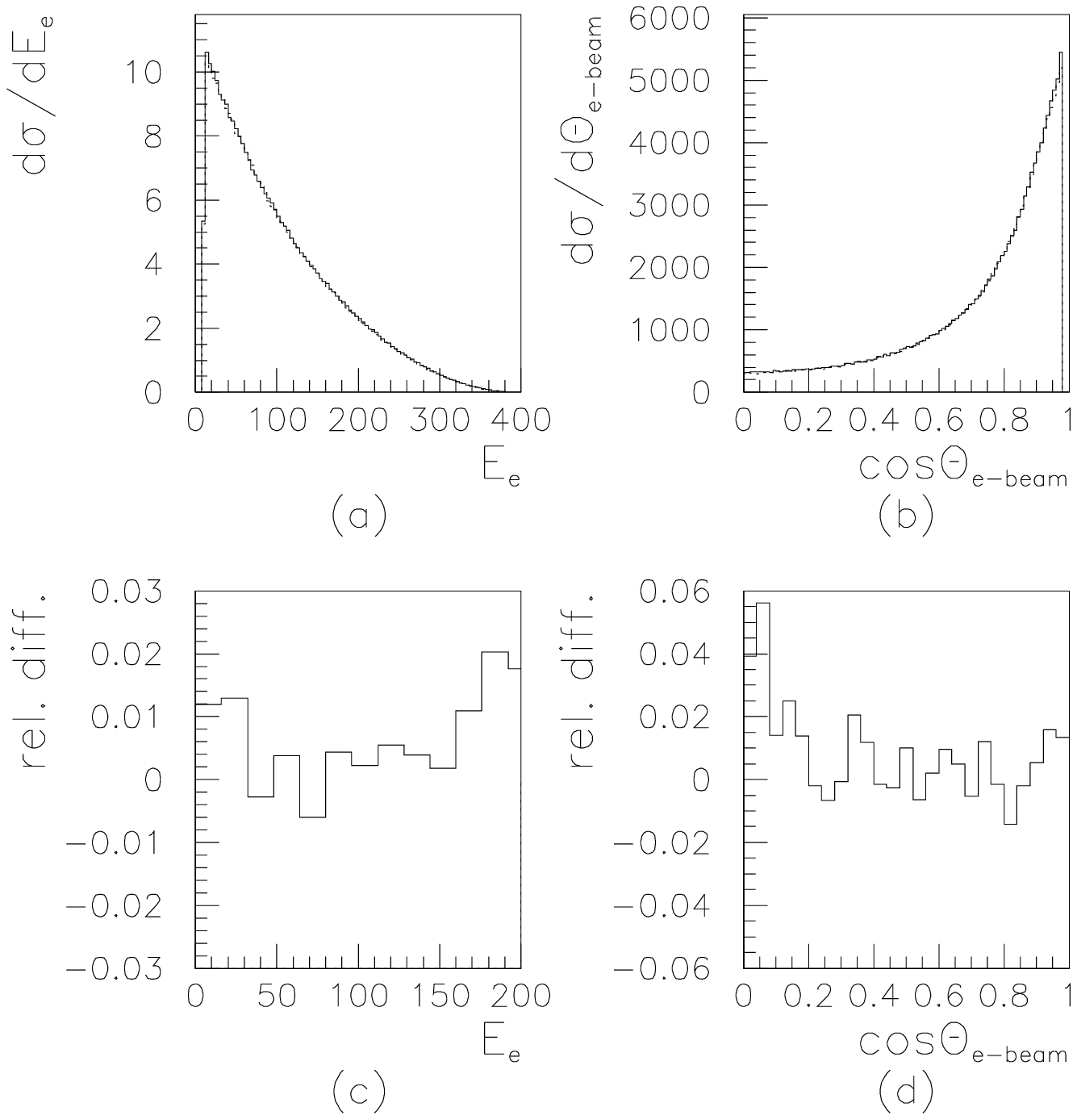}}
 \vspace*{-0.8cm}
\caption{ {\em As with the previous figure but for a centre-of-mass energy of 800GeV and 
with both photons being right-handed } $\sigma_{++}$.
\label{figdis800} }
\end{center}
\end{figure*}

On the other hand one should not expect to have $\sigma_{++}=\sigma_{--}$  when cuts 
are introduced. Indeed, consider the case of the narrow width approximation with full spin 
correlations. For like-sign photon helicities we remarked that only same-helicity $W$'s were 
produced. Moreover, the overwhelming contributions were those with transverse $W$'s whereby each 
photon transfers its helicity to the $W$. Before cuts are introduced $\sigma_{++}=\sigma_{--}$.
However, because of the chiral structure of the $W$ couplings, the electron is emitted in a 
preferential direction depending on the helicity of the $W$. Indeed, taking the $W^-$ flight
direction as a reference axis, in its rest frame 
a left-handed $W^-$ 
emits an electron in the forward direction (see~\ref{dfunctions})  
 while for a right-handed $W^-$ the electron is backward. At the energies we are
considering both $W$'s are very forward/backward (in the laboratory system)
moreover the boost  for the electron is substantial. This has the effect that for a 
left-handed $W^-$ the electron is emitted very forward in the lab system, but when an electron from
a right-handed $W^-$ is  boosted it tends to be further away 
from the beam  than is 
the electron from a left-handed electron $W^-$. As a result, this means that if one only cuts 
on forward-backward electrons (with respect to the beam) 
the cross section with left-handed photons $\sigma_{--}$ will be smaller than that with right-handed 
photons $\sigma_{++}$
. This has been explicitly checked by our calculations.
In the tables one clearly 
sees that the argument we have given above leading to 
 $\sigma_{++}>\sigma_{--}$ is explicitly confirmed both at 400 and 800~GeV,  
but seems to fail  for  $\sqrt{s_{\gamma \gamma}}=1600$~GeV. 
The reason the above argument  does not  lead to $\sigma_{++}>\sigma_{--}$ at this 
very high energy has to do with the fact that the other cuts 
we have taken,  more specifically
cuts on  the energy of the electron
and on the angles of the quarks/antiquarks, become very restrictive at this energy with an 
effect that counterbalances and washes out that of the simple description in terms of the angular cut on the 
electron. 
For instance, take the cut on the energy of the electron, $E_e> .0125 \sqrt{s}$, which is directly 
related to the angle $\theta^*_e$ 
that 
defines the direction of the electron in the rest frame of the $W$
(see Eq.~\ref{dfunctions}~-\ref{thetastar}). All events pass this cut at 400GeV and practically all pass 
this cut
at 800GeV, but not at 1600GeV.  However, events that are rejected by this cut, effectively on 
 $\theta^*_e$, correspond to electrons  that are backward (in the rest frame of the $W$) 
 therefore it is the electrons that 
originate from a right-handed $W^-$ that will be penalised. These  were the ones that, 
after boosting, passed the cuts of forward/backward electrons (with respect to the photon beam),
{\it i.e.} our previous argument becomes ineffective at $1600$GeV. 
\footnote{Fig.~\ref{figdis400} and Fig.~\ref{figdis800} show the drastic difference between the 
distributions in the energy of the electron produced in the $\sigma_{--}$ and $\sigma_{++}$ 
configurations. These distributions may be easily understood by the argument based on the 
$\cos \theta_e^*$.}
In fact, it turns out that 
the angular cut on the jets is even more penalising. 
First, if only this cut is imposed and since it is symmetric for the quark and the antiquark 
then $\sigma_{++}=\sigma_{--}$ for all energies. 
However, at very high energies the boost of the fermions becomes so large 
 that the angular cut on the {\em jets} 
that we have applied is tantamount to having imposed the same 
 angular cut on the $W^+$, {\it i.e} $\theta$, 
 and therefore on the
$W^-$ angle.   Consequently, the cut on the angle of the electron with the beam 
becomes redundant and does not cut many of the electrons from a left-handed 
$W^-$.  This has also been checked explicitly by our calculations.

From the tables one sees that, for all energies, for all combinations of photon helicities and
independently of the invariant mass cuts, the results are not very sensitive to how the width 
has been implemented on the all four-fermion final state. 
The discrepancies if any, are below the
per-mil level and thus consistent with our MC error.
A similar conclusion was reached in the
case of \eewwt so long as extreme forward electrons were rejected (this almost eliminates the 
$t$-channel photon exchange)\cite{wwsmlep2}. For the rest of the discussion, we will therefore 
compare the results 
of the improved narrow width approximation, those of keeping the ``resonant subset" 
(\underline{off-shell} $WW$
cross section Eq.~\ref{offshellcross}) with what one obtains with the full set of diagrams. 
As we discussed earlier, this ``resonant subset"
is expected to violate gauge invariance. We find that when no cuts on the invariant masses are imposed
the results one obtains with this subset are not to be trusted for any configuration of the 
photon helicities, especially as  the energy increases. At 400~GeV the discrepancy with the exact
result is about $3\%$ and increases to $17\%$ (!) at 1600~GeV for $\sigma_{++}$. 
It is only by imposing  
cuts on {\em both} leptonic and hadronic invariant masses that the ``resonant" subset 
(even for 1600~GeV) reproduces the 
complete calculation and this even when the invariant mass cut is 18~GeV. However, as soon as one relaxes
the cut on one of the invariant masses, namely the $W^-$, even a strict cut on the jet system of 
5~GeV is not enough to make the results  in accord with those of the complete calculation.  As expected the
discrepancy grows with the centre-of-mass energy and as the invariant mass cut gets loose. This is
also a manifestation of how a loss of gauge invariance can translate into a loss of unitarity. 

The narrow width approximation with full spin correlation taking into account the smearing factor fares 
much better than the ``resonant subset" (Eq.~\ref{offshellcross}) and turns out to be a good  
approximation.  
With a 5~GeV cut on both the leptons and the jets the narrow width approximation
reproduces the results of the complete calculation within the MC errors for all combinations of
helicities and for all energies. When the cut is relaxed to 18~GeV, the discrepancy is never 
larger than 3~per-mil and is within the errors for certain combinations of the photon helicities. 
If a cut of 18~GeV on the hadronic system  only is imposed the agreement worsens, and for the lower
energies it affects the different configurations of the photon helicities differently. 
For instance, with our set of cuts, at $400$~GeV $\sigma_{--}$ is reproduced at the per-mil level 
by the improved narrow width approximation while $\sigma_{++}$ and the $J_Z=2$ are off by one
per-cent.  On the other hand, at $800$GeV while the agreement remains sensibly the same for 
$\sigma_{++}$ and very slighlty worsens  in the $J_Z=2$ 
to be $\simeq 1.3\%$, for $\sigma_{--}$ one has agreement at $\simeq 1.5\%$. At 1600GeV the
agreement for $\sigma_{++}$ is below $2\%$ while for the other photon helicities it reaches $3\%$.
At this particular energy, one may thus have to revert to the result based on the complete set of 
diagrams. \\
\noindent If one would like to keep more statistics by not imposing an invariant mass cut
neither on the hadronic
nor the leptonic system, 
the narrow width approximation generally does not describe the results of the 
4-fermion final 
state, especially as the energy increases. It is only with the particular helicity 
$\sigma_{--}$ at 400~GeV that the agreement is within the per-mil, otherwise the 
discrepancy at 
400 and 800~GeV is about $1-2\%$ reaching $4-5\%$ at 1600~GeV. Nonetheless, 
 these results are clearly far better 
than the ones obtained with the off-shell $WW$ cross section (``resonant" subset). \\

Another important issue to check is whether our improved narrow width 
approximation reproduces 
the various distributions. For instance one may wonder whether the excellent 
agreement for the 
$\sigma_{--}$ at 400GeV at the level of the integrated cross section is 
fortuitous. That is, does  the approximation reproduce the various distributions 
one obtains with the full set of diagrams
as well as it reproduces the integrated cross section? or could 
the distributions be different and yet agree when integrated over 
all the events? 
To analyse this issue,  we have 
looked at two
distributions, the energy of the electron and the angle of the electron with 
the beam. Note that for
the latter, non resonant diagrams whereby a photon splits into an 
electron/positron pair can contribute 
substantially when the electron is very forward. Therefore this distribution could 
indicate if the cut on the forward electrons that we have imposed could be made stricter 
in order to reach a better agreement with the ``exact" result.\\
\noindent To illustrate our 
point we restrict ourselves to 
the case $\sigma_{--}$ at 400GeV and $\sigma_{++}$ at 800GeV, and only study the realistic case 
where a cut of 
18GeV is imposed on the jets only. At 400GeV one sees that both the angular 
and the energy 
distributions are reproduced extremely well (see Fig.~\ref{figdis400}). 
In fact comparing the distributions, side by side, one can hardly see the 
discrepancies. To better quantify the agreement we 
looked at the 
relative difference between the two distributions. 
For the energy distribution the agreement is 
always below the per-cent (the
peaks one observes in the bins at the lower and upper ends have very 
low statistics and are not to
be trusted). For the angular distribution, the agreement fluctuates 
between $\sim -2\%$ and $2\%$ in some
of the bins, but for $|\cos\theta_{e \gamma}|>.8$ where the distributions peak
and where the events are concentrated the agreement is much better.
One can make the same statement about the distributions at 800GeV (see Fig.~\ref{figdis800}). 
The distributions are reproduced quite well and the agreement 
is of the same order as the one reached for the integrated cross section, a slightly 
less good agreement is observed only in regions that are much less populated. For instance,  
the relative difference in the angular distribution of the electron is maximum for 
events in the central region.

One can therefore
conclude by saying that as long as moderate cuts on the di-jet invariant mass 
are imposed, the
narrow width approximation including full spin correlations and smearing 
reproduces the results of
the complete $ l \nu_l j j$ channel at the per-cent level for energies
 up to 1~TeV, and even much 
better at energies around 400GeV. Above the TeV, 
unless much stricter invariant mass cuts are imposed (meaning a loss of
 statistics), for precision
measurements one needs to rely on the result of the complete calculation
 based on the full set of
diagrams describing the 4-fermion final states. The latter computation, 
with the
requirement that one keeps particles within the central region, is not sensitive to 
the  different implementations of the $W$
width. We also find that when the agreement at the level of the integrated 
cross section is obtained
the distributions are  reproduced with the same accuracy especially 
in regions where one has the bulk of the events. \\
\noindent To simulate a more realistic $\gamma \gamma$ set-up one should have 
convoluted our cross sections with polarised luminosity functions that describe the 
energy spectrum of the photons. This could be very easily implemented 
especially with the improved narrow width  approximation. Most of the analyses 
on the laser induced \gag physics have taken an ideal spectrum based on a theoretical 
calculation of this spectrum\cite{PhotonCol} 
that does not simulate the full conditions of the conversion of an electron beam 
by an intense laser. More realistic simulations of the spectrum have only very recently 
been considered\cite{Sheffieldtelnov}. Because of the uncertainty in these spectra we have preferred
not to consider any particular spectrum especially that our principal aim was to provide a 
calculation for four-fermion 
final states in polarised \gag reactions and to enquire whether an approximation based on the resonant 
$WW$ diagrams would be sufficient for future analyses.

\end{document}